\documentclass[twocolumn]{aastex631}

\usepackage[caption = false]{subfig}
\usepackage{csquotes}

\usepackage[normalem]{ulem}
\usepackage{xcolor}

\usepackage{mathtools}

\begin{document}

\title{Nature of Transonic Sub-Alfvénic Turbulence and Density Fluctuations in the Near-Sun Solar Wind: Insights from Magnetohydrodynamic Simulations and Nearly-Incompressible Models}

\author{Giuseppe Arrò}
\affiliation{Department of Physics, University of Wisconsin-Madison, Madison, WI 53706, USA}
\affiliation{Los Alamos National Laboratory, Los Alamos, NM 87545, USA}

\author{Hui Li}
\affiliation{Los Alamos National Laboratory, Los Alamos, NM 87545, USA}

\author{Gary P. Zank}
\affiliation{Department of Space Science, University of Alabama in Huntsville, Huntsville, AL 35805, USA}
\affiliation{Center for Space Plasma and Aeronomic Research (CSPAR), University of Alabama in Huntsville, Huntsville, AL 35805, USA}

\author{Lingling Zhao}
\affiliation{Department of Space Science, University of Alabama in Huntsville, Huntsville, AL 35805, USA}
\affiliation{Center for Space Plasma and Aeronomic Research (CSPAR), University of Alabama in Huntsville, Huntsville, AL 35805, USA}

\author{Laxman Adhikari}
\affiliation{Department of Space Science, University of Alabama in Huntsville, Huntsville, AL 35805, USA}
\affiliation{Center for Space Plasma and Aeronomic Research (CSPAR), University of Alabama in Huntsville, Huntsville, AL 35805, USA}

\begin{abstract}

Recent Parker Solar Probe measurements have revealed that solar wind (SW) turbulence transits from a subsonic to a transonic regime near the Sun, while remaining sub-Alfvénic. These observations call for a revision of existing SW models, where turbulence is considered to be both subsonic and sub-Alfvénic. In this Letter, we introduce a new magnetohydrodynamic (MHD) model of Transonic sub-Alfvénic Turbulence (TsAT). Our model shows that turbulence is effectively nearly-incompressible (NI) and has a \enquote{2D + slab} geometry not only in the subsonic limit, but also in the transonic regime, as long as it remains sub-Alfvénic, a condition essentially enforced everywhere in the heliosphere by the strong local magnetic field. These predictions are consistent with 3D MHD simulations, showing that transonic turbulence is dominated by low frequency quasi-2D incompressible structures, while compressible fluctuations are a minor component corresponding to low frequency slow modes and high frequency fast modes. Our new TsAT model extends existing NI theories of turbulence, and is potentially relevant for the theoretical and numerical modeling of space and astrophysical plasmas, including the near-Sun SW, the solar corona, and the interstellar medium.

\end{abstract}


\section{Introduction} 

Satellite measurements near 1 au from the Sun and beyond have revealed that solar wind (SW) turbulence is weakly compressible, with density fluctuations that are typically less than $10\%$ of the background density, and subsonic velocity fluctuations $\delta u$, with turbulent sonic Mach numbers $M_S\!=\! \delta u/c_S\!\lesssim\!0.1$, where $c_S$ is the sound speed \citep{roberts1987nature,montgomery1987density,matthaeus1990evidence,matthaeus1991nearly}. These observations stimulated the development of nearly-incompressible (NI) models of SW turbulence, based on the $M_S\!\ll\!1$ assumption. In the $M_S\!\ll\!1$ regime, compressible magnetohydrodynamic (MHD) equations can be expanded about a low frequency incompressible state, with small amplitude high frequency compressible corrections quasi-linearly coupled to the dominant incompressible flow \citep{matthaeus1988nearly,matthaeus1991nearly}.

The SW dynamics is also affected by the strong interplanetary magnetic field, making turbulence anisotropic \citep{shebalin1983anisotropy,horbury2008anisotropic,wicks2010power,adhikari2024mhd}, and sub-Alfvénic, with turbulent Alfvén Mach numbers $M_A\!=\!\delta u/c_A\!\ll\!1$, where $c_A$ is the Alfvén speed. Early NI models assumed $M_A\!\sim\!\mathcal{O}(1)$, but extensions were developed, incorporating the $M_A\!\ll\!1$ condition, together with $M_S\!\ll\!1$, and varying plasma beta $\beta\!=\!2\,c_S^2/c_A^2$ \citep{zank1992waves,zank1993nearly,bhattacharjee1998weakly,hunana2010inhomogeneous,zank2017theory}. The main conclusion drawn from these NI models is that subsonic sub-Alfvénic turbulence exhibits a \enquote{2D + slab} structure, meaning that turbulent fluctuations consist of a majority low frequency 2D incompressible component, plus a minority of high frequency 3D waves. 

The NI turbulent regime described so far potentially breaks down near the Sun, where turbulent velocity fluctuations are expected to increase in amplitude \citep{cranmer2017origins,adhikari2020solar,adhikari2022modeling}, and $\beta$ is small \citep{kasper2021parker}, implying $c_S\!\ll\!c_A$. This results in a larger $M_S$, suggesting that turbulence might become more compressible near the Sun. Indeed, Parker Solar Probe (PSP) measurements have recently shown that SW turbulence becomes transonic at about $11\,R_\odot$ (solar radii) from the Sun, where $M_S\!\sim\!1$ and $M_A\!\ll\!1$ \citep{zhao2025transonic}. This transition from a subsonic to a transonic turbulent regime is accompanied by a significant increase in density fluctuations, up to more than $20\%$ of the background density. 

Recent MHD simulations of subsonic turbulence have revealed that turbulent fluctuations mainly consist of low frequency quasi-2D modes, rather than waves \citep{gan2022existence,fu2022nature,arro2025}. Low frequency fluctuations are typically interpreted as coherent structures \citep{karimabadi2013coherent,papini2021spacetime,arro2023generation,arro2024large,Espinoza-Troni_2025, zhao2025non}, produced by an inverse energy cascade \citep{arro2025}. The prevalence of low frequency quasi-2D fluctuations over Alfvén waves (AWs), slow modes (SMs), and fast modes (FMs) is consistent with NI models of turbulence \citep{zank1992waves,zank1993nearly,zank2017theory}, and with SW observations \citep{perrone2016compressive,roberts2017multipoint,zhao2023observations,zank2024characterization}. However, it is unclear whether this picture of turbulence continues to hold also in the newly observed $M_S\!\sim\!\mathcal{O}(1)$ near-Sun regime, and to our knowledge, no model of transonic SW turbulence has ever been developed. 

In this Letter, we study transonic sub-Alfvénic turbulence using a 3D MHD simulation initialized with typical near-Sun SW parameters. We find that even in the transonic regime, turbulence is dominated by low frequency quasi-2D incompressible structures, while compressible fluctuations are a minor component in the form of low frequency SMs and high frequency FMs. These results are consistent with a new MHD model of Transonic sub-Alfvénic Turbulence (TsAT) that we derive, showing that turbulence in the $M_S\!\sim\!\mathcal{O}(1)$, $M_A\!\ll\!1$ regime is effectively NI, with a 2D $+$ slab structure, similarly to subsonic turbulence.

\section{Simulation Setup}

We performed our simulation using the MHD code \textit{Athena++} \citep{stone2020athena++}. We consider a domain of size $L_z\!=\!3L_y\!=\!3L_x\!=\!6\,\pi$ (in arbitrary units $L_0$), sampled by a uniform periodic mesh with $256\!\times\!512^2$ points. The plasma is initially at rest, with uniform density $\rho_0$ and guide field $\textbf{B}_0\!=\!B_0\widehat{\textbf{z}}$. Pressure $P$ is isothermal, with $\beta\!=\!2\,c^2_S/c^2_A\!=\!0.045$. Turbulence is driven using the Langevin antenna method \citep{tenbarge2014oscillating}, with driving frequency $\omega_0\!=\!0.8\,\tau_A^{-1}$ and decorrelation rate $\gamma_0\!=\!-0.7\,\tau_A^{-1}$ (where $\tau_A\!=\!L_0/c_A$). Injected velocity and magnetic field fluctuations are incompressible and polarized in the plane perpendicular to $\textbf{B}_0$, with $1\!\leqslant\!k_{\parallel}L_z/2\pi\!\leqslant\!3$ and $1\!\leqslant\!k_{\perp}L_x/2\pi\!\leqslant\!4$, where $k_{\parallel}$ and $k_{\perp}$ are wavenumbers parallel and perpendicular to $\textbf{B}_0$. Viscous and resistive dissipation are included, with kinetic and magnetic Reynolds numbers $R_e\!=\!R_m\!=\!4000$. The continuous driving produces a quasi-stationary turbulent state after about $20\,\tau_A$, characterized by density, velocity and magnetic field fluctuations with rms amplitudes $\delta \rho_{rms}/\rho_0\!\simeq\!0.22$, $\delta u_{rms}/c_A\!=\!M_A\!\simeq\!0.155$ and $\delta B_{rms}/B_0\!\simeq\!0.153$. Turbulent fluctuations have $M_S\!=\!\delta u_{rms}/c_S\!\simeq\!1.03$, and cross helicity $\sigma_C\!=\!\bigl< 2 \, \textbf{u} \cdot \delta\textbf{B}/\sqrt{\rho} \bigr>/\bigl< u^2 + \delta B^2/\rho \bigr>\!\simeq\!0.64$, where $\delta \textbf{B}=\textbf{B}-\textbf{B}_0$, and $\bigl<\cdot\bigr>$ is the box average. These parameters are consistent with observations of near-Sun transonic sub-Alfvénic turbulence \citep{zhao2025transonic}.

\section{Results}

\begin{figure*}[t]
\centering
\subfloat{
\includegraphics[width=0.48\linewidth]{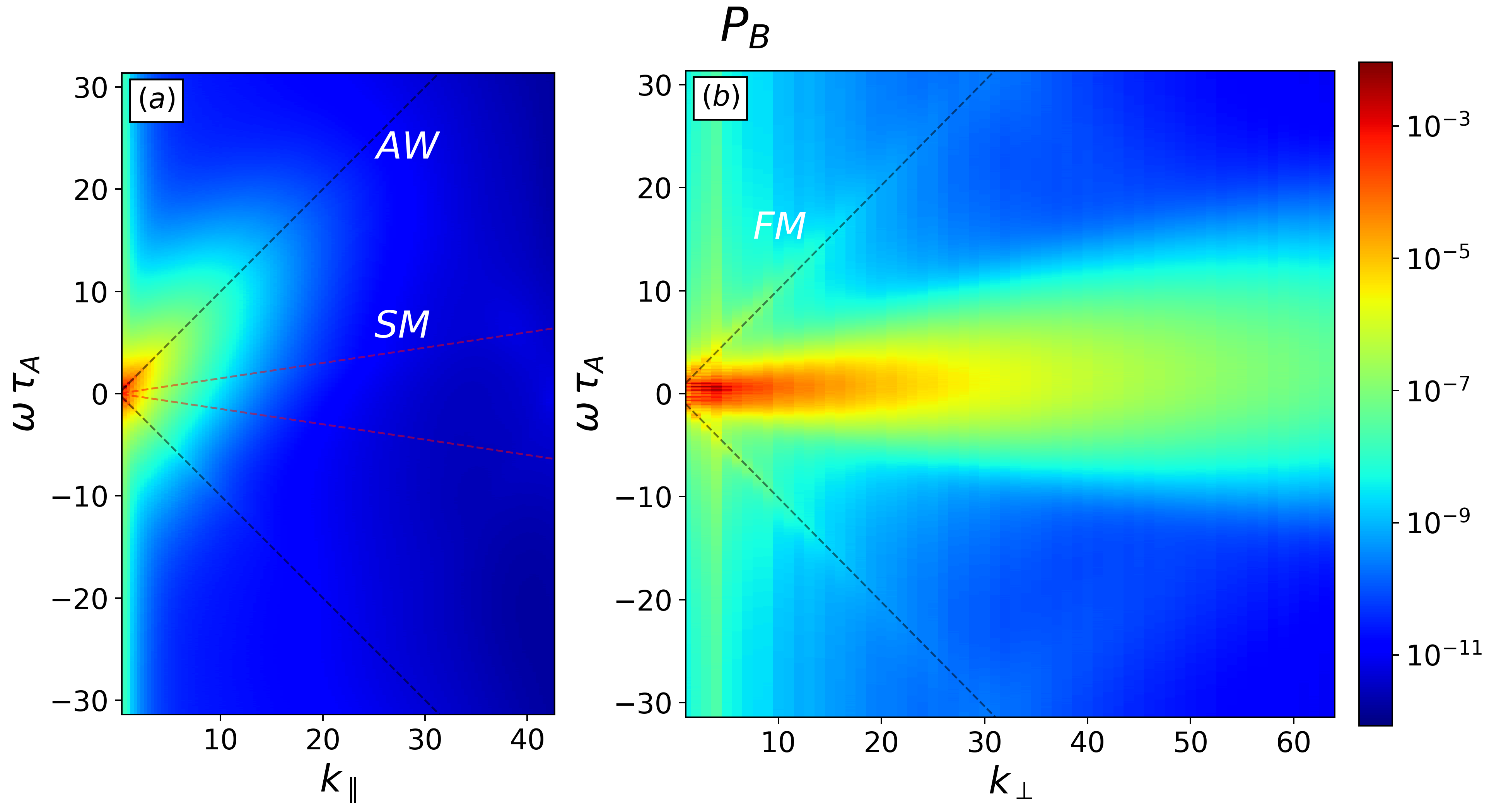}
}
\subfloat{
\includegraphics[width=0.48\linewidth]{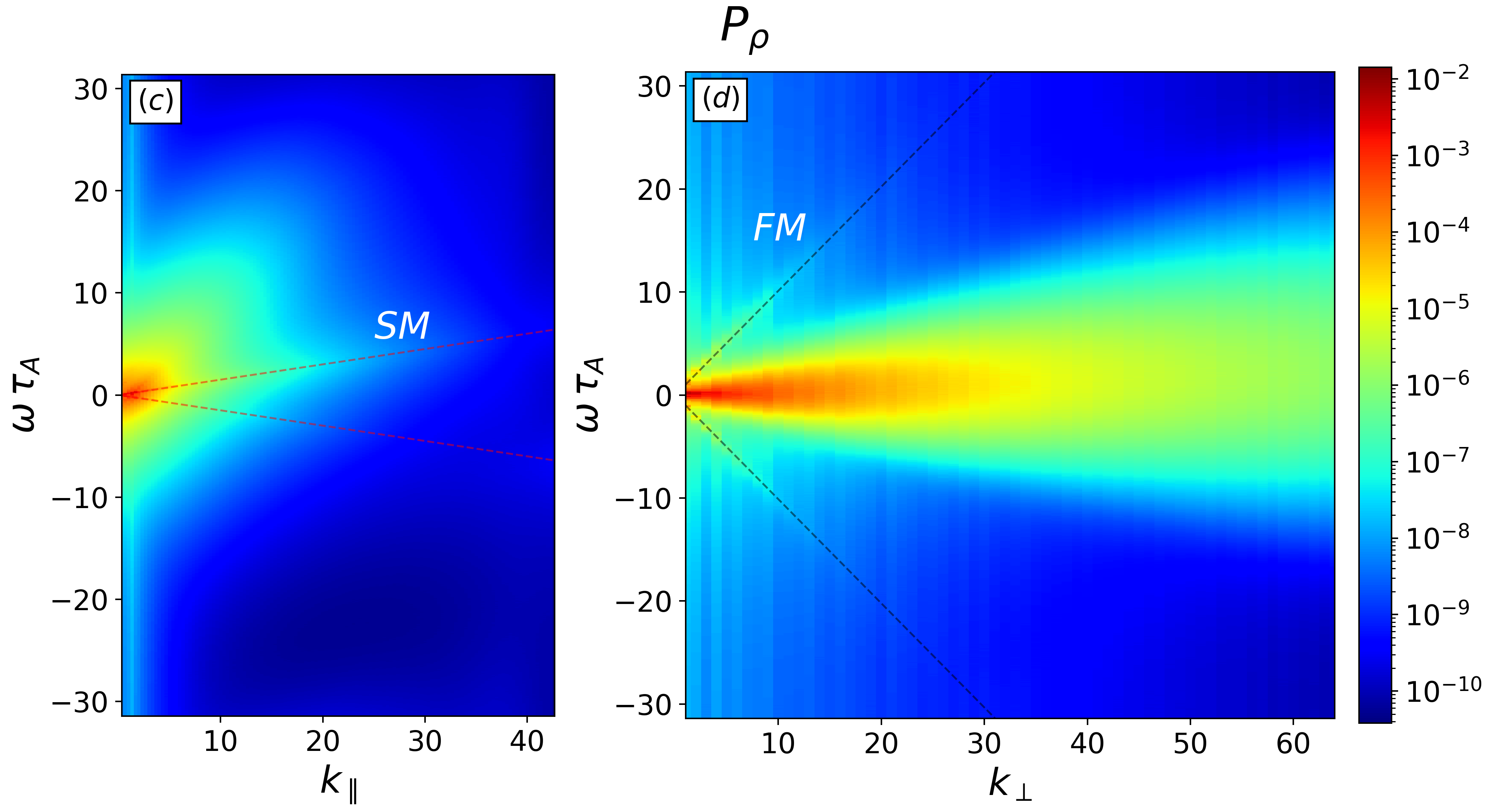}
}
\\
\subfloat{
\includegraphics[width=0.48\linewidth]{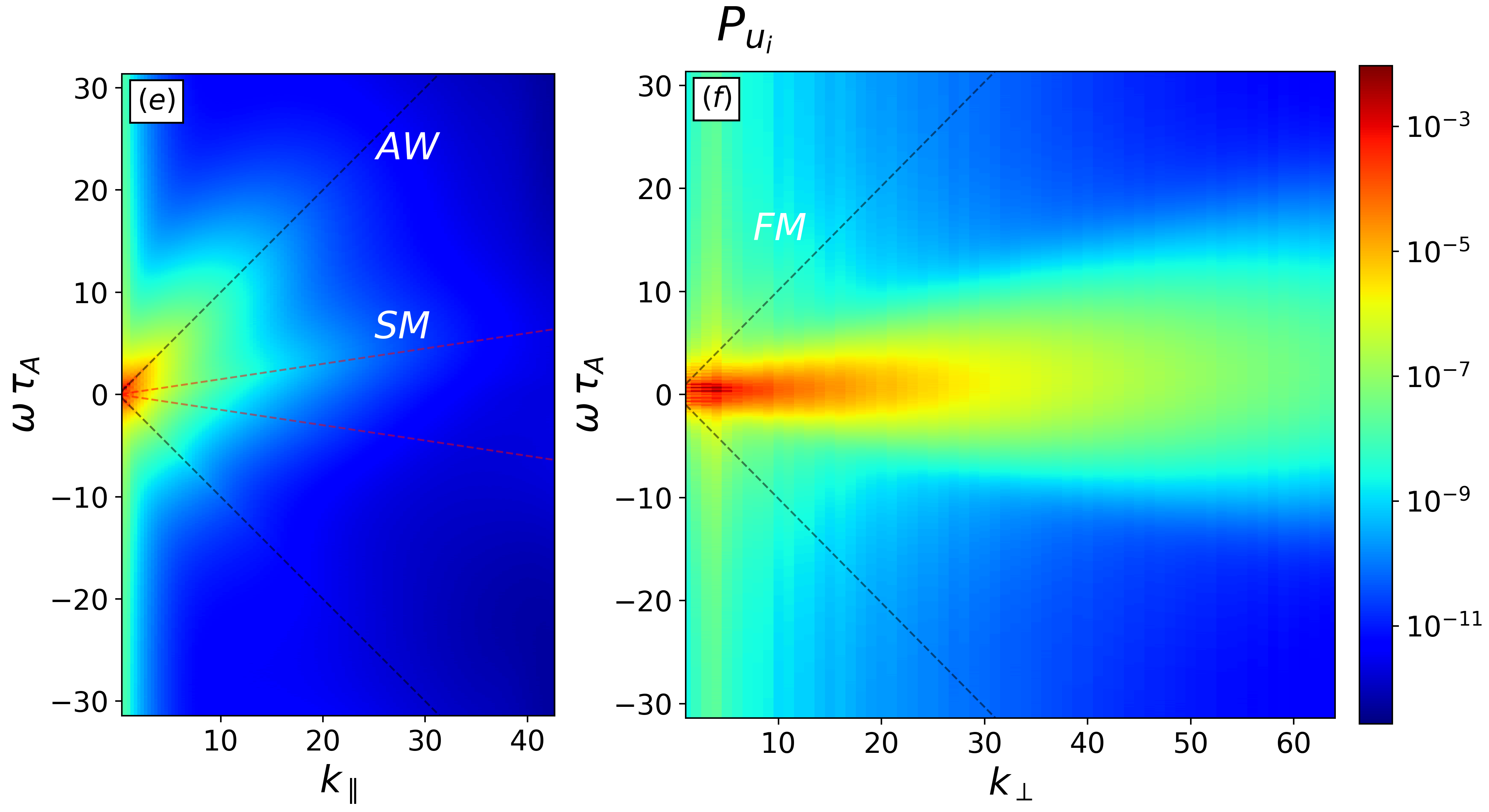}
}
\subfloat{
\includegraphics[width=0.48\linewidth]{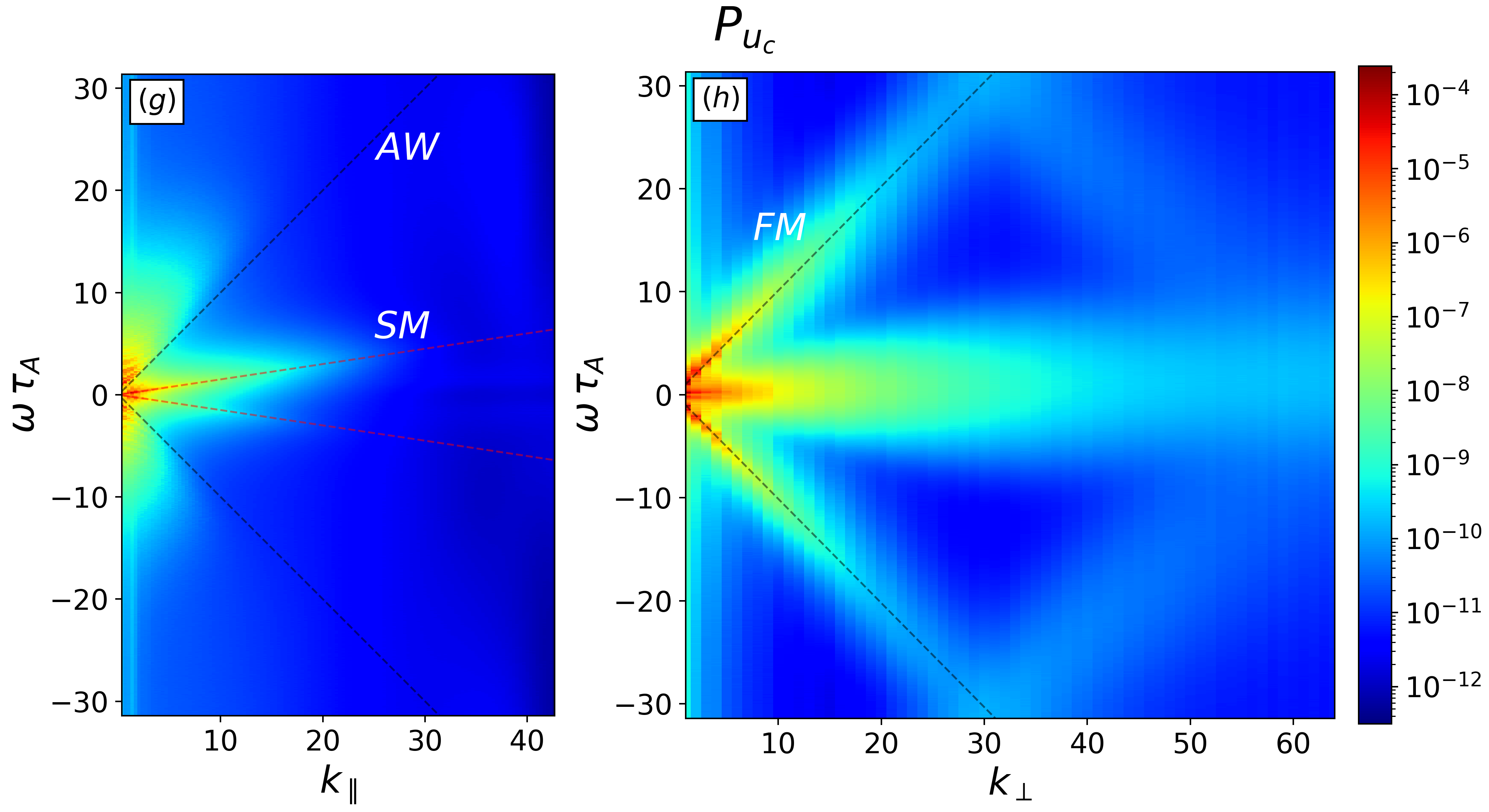}
}
\caption{$(k_{\parallel},\,\omega)$ and $(k_{\perp},\,\omega)$ projections of $P_B$ (a)-(b), $P_{\rho}$ (c)-(d), $P_{u_i}$ (e)-(f), and $P_{u_c}$ (g)-(h), with dashed lines indicating dispersion relations of Alfvén waves (AW) and slow modes (SM) for $k_{\perp}\!=\!0$, and fast modes (FM) for $k_{\parallel}\!=\!0$.}
\label{Prho}
\end{figure*}

In this section, we analyze the spectral properties of our turbulence simulation in wavenumber-frequency space, investigating the presence of waves and structures. Then, we derive a new MHD model of transonic sub-Alfvénic turbulence to interpret our numerical results. 

Figures~\ref{Prho}(a)-(b) show $(k_{\parallel},\,\omega)$ and $(k_{\perp},\,\omega)$ projections of the magnetic field wavenumber-frequency spectrum $P_{B}$ (where $\omega$ is the frequency), calculated over time interval $[20\,\tau_A,\,120\,\tau_A]$, with data sampled every $\Delta t\!=\!0.1\,\tau_A$. The two projections correspond to $P_{B}(k_{\parallel},\,\omega) = \int P_{B}(k_{\perp},\,k_{\parallel},\,\omega) \, dk_{\perp}$ and $P_{B}(k_{\perp},\,\omega) = \int P_{B}(k_{\perp},\,k_{\parallel},\,\omega) \, dk_{\parallel}$. Similarly to subsonic turbulence, we find that in the transonic regime most magnetic field fluctuations correspond to low $\omega$ modes, while AWs, SMs and FMs (dashed lines) are a minor component. Low $\omega$ fluctuations are highly anisotropic, with a much wider distribution in $k_{\perp}$ than in $k_{\parallel}$, implying a quasi-2D geometry. Since low $\omega$ fluctuations do not follow the dispersion relations of waves, we dub them non-wave modes (NWMs). $P_{B}(k_{\parallel},\,\omega)$ is skewed toward positive $\omega$ because of the high $\sigma_C$, making turbulence imbalanced \citep{lugones2019spatio,arro2025}. Figures~\ref{Prho}(c)-(d) show $(k_{\parallel},\,\omega)$ and $(k_{\perp},\,\omega)$ projections of the density wavenumber-frequency spectrum $P_{\rho}$. We see that even density fluctuations mainly consist of low $\omega$ quasi-2D NWMs, dominating over SMs and FMs. To understand whether NWMs are incompressible or compressible, we analyze the velocity wavenumber-frequency spectrum, separating the velocity field $\textbf{u}$ into its incompressible and compressible components, $\textbf{u}_i$ and $\textbf{u}_c$, with $\nabla\cdot\textbf{u}_i\!=\!0$ and $\nabla\times\textbf{u}_c\!=\!0$ \citep{bhatia2012helmholtz}. The ratio between the rms amplitudes of $\textbf{u}_c$ and $\textbf{u}_i$ is $\delta u_{c,rms}/\delta u_{i,rms}\!\simeq\!0.15$, meaning that incompressible fluctuations are globally stronger than compressible ones. Figures~\ref{Prho}(e)-(f) show $(k_{\parallel},\,\omega)$ and $(k_{\perp},\,\omega)$ projections of the incompressible velocity spectrum $P_{u_i}$, while panels (g)-(h) show the corresponding projections of the compressible velocity spectrum $P_{u_c}$. We find that $\textbf{u}_i$ is almost entirely made up of NWMs, with a small contribution from AWs, and little to no energy associated with SMs and FMs. On the other hand, $\textbf{u}_c$ essentially consists of SMs and FMs. The $(k_{\parallel},\,\omega)$ projection of $P_{u_c}$, panel (g), reveals the presence of SMs at low $\omega$, and a regular wavy pattern extending toward higher $\omega$, corresponding to the projection of FMs in the $(k_{\parallel},\,\omega)$ plane. The $(k_{\perp},\,\omega)$ projection of $P_{u_c}$, panel (h), shows the presence of high frequency FMs, and some energy distributed around low $\omega$, corresponding to the projection of SMs in the $(k_{\perp},\,\omega)$ plane. Low frequency SMs extend over a much narrower $k_{\perp}$ range and are much weaker in amplitude than incompressible NWMs observed over the same frequency range in $P_{u_i}$. Therefore, we find that NWMs are incompressible and account for the majority of magnetic and density fluctuations. 

\begin{figure*}[t]
\centering
\subfloat{
\includegraphics[width=0.32\linewidth]{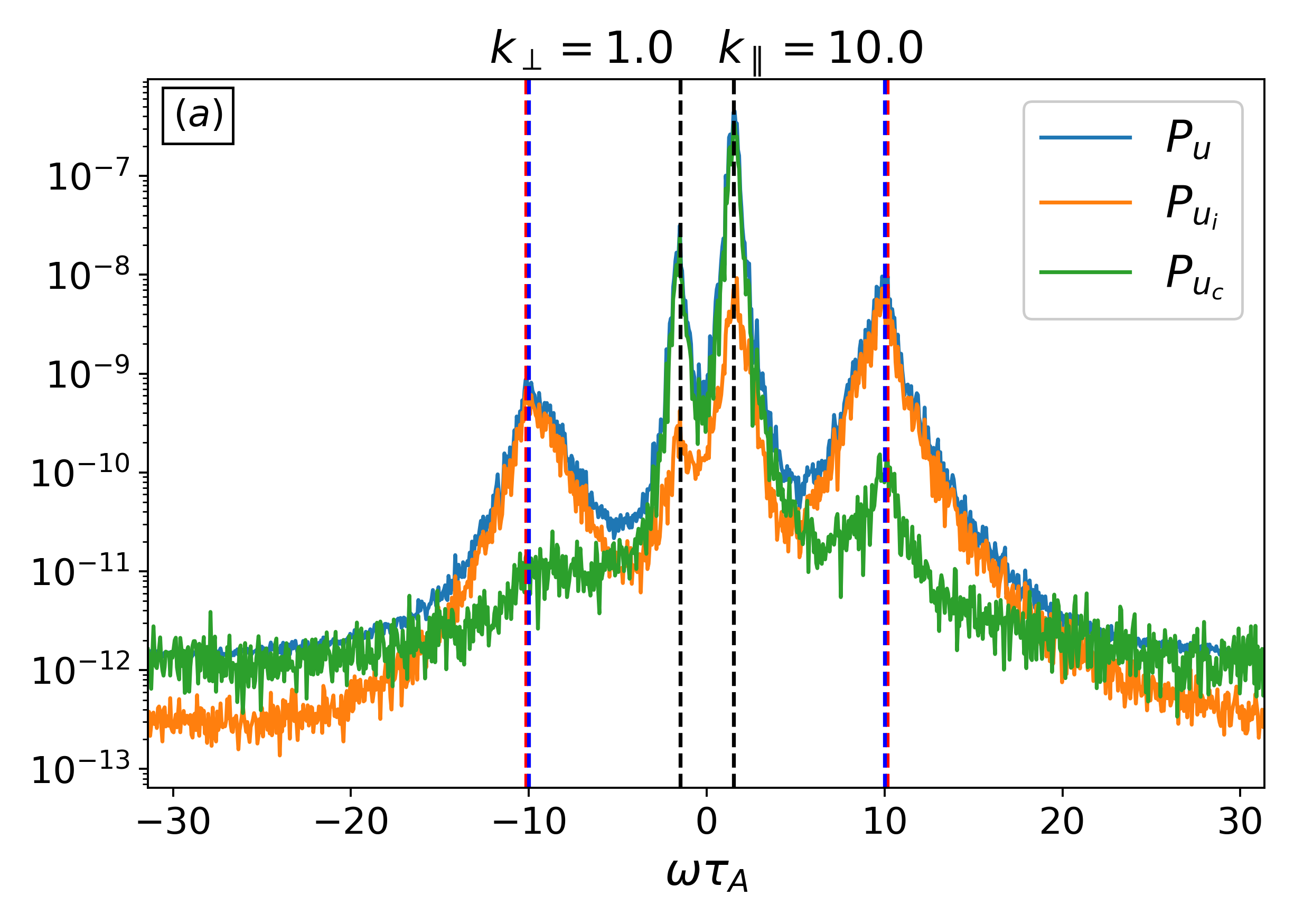}
}
\subfloat{
\includegraphics[width=0.32\linewidth]{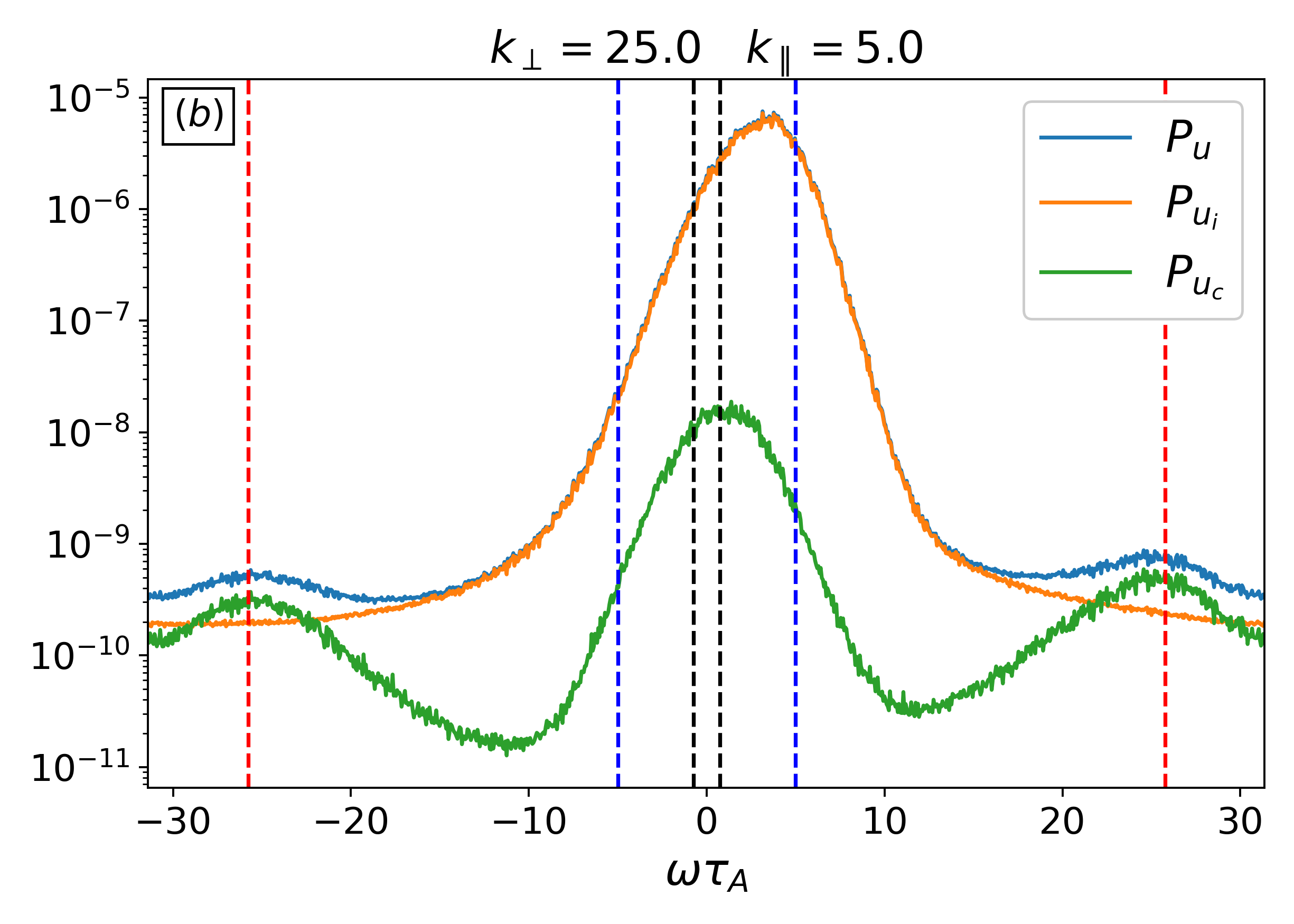}
}
\subfloat{
\includegraphics[width=0.32\linewidth]{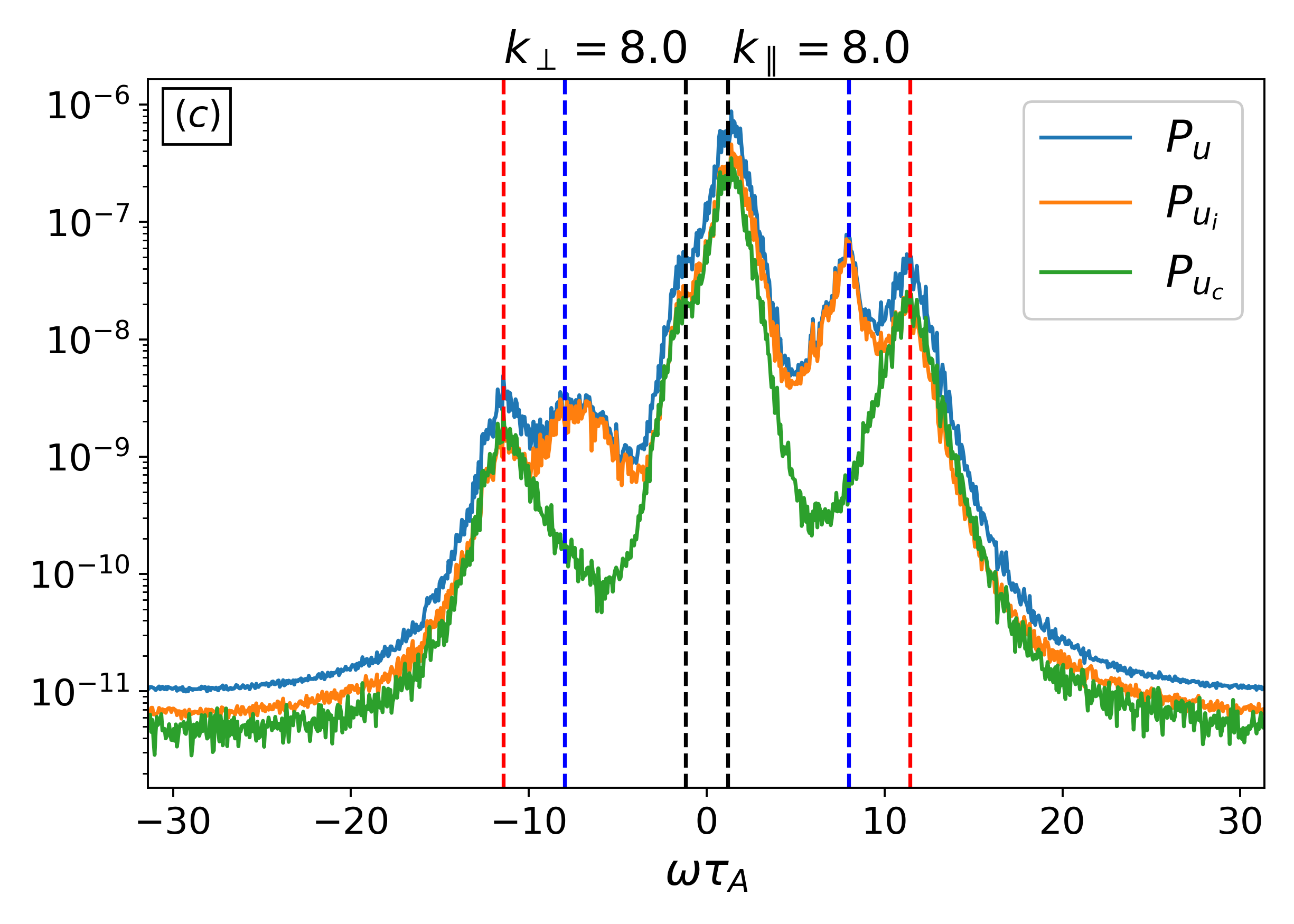}
}
\caption{Frequency spectra of $\textbf{u}$, $\textbf{u}_i$ and $\textbf{u}_c$ at $(k_{\perp}\!=\!1,\,k_{\parallel}\!=\!10)$ (a), $(k_{\perp}\!=\!25,\,k_{\parallel}\!=\!5)$ (b), and $(k_{\perp}\!=\!8,\,k_{\parallel}\!=\!8)$ (c), with vertical dashed lines indicating the corresponding frequencies of Alfvén waves (blue), slow modes (black), and fast modes (red).}
\label{TMHD}
\end{figure*}

To further assess the relative importance of waves with respect to NWMs, we compare frequency spectra of $\textbf{u}$, $\textbf{u}_i$ and $\textbf{u}_c$ at different propagation directions. Figure~\ref{TMHD}(a) shows the total velocity spectrum $P_u$, together with $P_{u_i}$ and $P_{u_c}$, as functions of $\omega$, for quasi-parallel modes with $(k_{\perp}\!=\!1,\,k_{\parallel}\!=\!10)$. We see that $P_u$ has four peaks, two of which correspond to forward and backward propagating low frequency SMs (black vertical dashed lines), with $P_u\!\simeq\!P_{u_c}$, while the other two smaller peaks correspond to a superposition of high frequency FMs and AWs (red and blue vertical dashed lines), with $P_u\!\simeq\!P_{u_i}$. Thus, parallel propagating fluctuations are dominated by low $\omega$ compressible SMs, with a smaller high $\omega$ incompressible contribution from FMs and AWs. In Figure~\ref{TMHD}(b) we compare $P_u$, $P_{u_i}$ and $P_{u_c}$ for quasi-perpendicular modes with $(k_{\perp}\!=\!25,\,k_{\parallel}\!=\!5)$. In this case, $P_u\!\simeq\!P_{u_i}$, with a low $\omega$ peak that does not match the frequency of waves, and with negligible high $\omega$ compressible contributions from FMs. The largest peak corresponds to incompressible NWMs, representing the majority of perpendicular fluctuations. Finally, Figure~\ref{TMHD}(c) shows $P_u$, $P_{u_i}$ and $P_{u_c}$ for oblique modes with $(k_{\perp}\!=\!8,\,k_{\parallel}\!=\!8)$. At this propagation direction, low frequencies are dominated by SMs, with $P_{u_i}\!\simeq\!P_{u_c}$, while higher frequencies show a significant contribution from AWs, with $P_u\!\simeq\!P_{u_i}$, and FMs, with $P_{u_i}\!\simeq\!P_{u_c}$. Hence, modes with $k_{\perp}\!\gg\!k_{\parallel}$ mainly consist of incompressible NWMs, while parallel and oblique fluctuations are essentially low frequency SMs and high frequency AWs and FMs. 

Overall, we find that transonic sub-Alfvénic turbulence is mostly incompressible, with low frequency quasi-2D NWMs dominating over waves. This seemingly counterintuitive behavior can be understood by taking the $M_S\!\sim\!\mathcal{O}(1)$, $M_A\!\ll\!1$ limit of compressible MHD equations. Following the method outlined in \citet{zank1993nearly}, we consider the normalized MHD equations 
\begin{equation}
\begin{gathered}
\bigl[\partial_t + (\textbf{u}\cdot \nabla)\bigr] \rho = -\rho(\nabla\cdot\textbf{u}),
\end{gathered}    
\label{density}
\end{equation}

\begin{equation}
\begin{gathered}
\rho \bigl[ \partial_t  + \bigl( \textbf{u} \cdot \nabla \bigr) \bigr] \textbf{u} = - \frac{\nabla P}{M^2_S} + \frac{1}{M^2_A} \bigl( \nabla \times \textbf{B} \bigr) \times \textbf{B},
\end{gathered}
\label{momentum}
\end{equation}

\begin{equation}
\begin{gathered}
\bigl[\partial_t + (\textbf{u}\cdot \nabla)\bigr] P = -\gamma P (\nabla\cdot\textbf{u}),
\end{gathered}
\label{pressure}
\end{equation}

\begin{equation}
\begin{gathered}
\bigl[\partial_t + (\textbf{u}\cdot \nabla)\bigr] \textbf{B} = (\textbf{B}\cdot\nabla) \textbf{u} - \textbf{B} (\nabla\cdot\textbf{u}),
\end{gathered}
\label{induction}
\end{equation}
with $\gamma$ being the polytropic index. In looking for turbulent solutions to these equations, three time scales are identified, namely $T_T\!=\!\lambda/U_0$, associated with the convective turbulent dynamics (where $\lambda$ is a characteristic length scale of turbulence), the acoustic time scale $T_S\!=\!\lambda/c_S$, and the Alfvénic time scale $T_A\!=\!\lambda/c_A$. In the $M_S\!\sim\!\mathcal{O}(1)$, $M_A\!=\!\epsilon\!\ll\!1$ regime, equation (\ref{momentum}) becomes
\begin{equation}
\begin{gathered}
\rho \bigl[ \partial_t  + \bigl( \textbf{u} \cdot \nabla \bigr) \bigr] \textbf{u} = - \nabla P + \frac{1}{\epsilon^2} \bigl( \nabla \times \textbf{B} \bigr) \times \textbf{B},
\end{gathered}
\label{Tmomentum}
\end{equation}
with $T_T/T_S\!\sim\!1/M_S\!\sim\!\mathcal{O}(1)$ and $T_T/T_A\!\sim\!1/M_A\!\gg\!1$, implying that turbulence and SMs evolve on similar slow time scales, while AWs and FMs are much faster. With such time scale separation, we can assume that turbulent solutions consist of two components, a majority low frequency turbulent state, and a minority component containing high frequency fluctuations. Following this ansatz, the strategy is finding a low frequency solution to MHD equations in the $\epsilon\!\rightarrow\!0$ limit, and then introduce small amplitude corrections, building an approximated solution valid for small but finite $\epsilon$. The majority low frequency component, labeled as \enquote{$\infty$}, can be obtained using Kreiss's theorem \citep{kreiss1980problems}, stating that time derivatives of low frequency solutions must be independent of $\epsilon$. We thus consider a low frequency solution of the form
\begin{equation}
\begin{gathered}
\rho = \rho_0 + \epsilon \rho^{\infty},
\quad
\textbf{u} = \textbf{u}^{\infty},
\\[10pt]
P = P_0 + \epsilon P^{\infty},
\quad
\textbf{B} = \textbf{B}_0 + \epsilon \textbf{B}^{\infty},
\end{gathered}
\end{equation}
where $\rho_0$, $P_0$ and $\textbf{B}_0\!=\!B_0\widehat{\textbf{z}}$ are constants. Starting from $\rho$, its time derivative gives
\begin{equation}
\begin{gathered}
\partial_t \rho^{\infty} = - \nabla \cdot \bigl( \rho^{\infty}  \textbf{u}^{\infty} \bigr) - \frac{\rho_0}{\epsilon} \bigl( \nabla \cdot \textbf{u}^{\infty} \bigr).
\end{gathered}    
\end{equation}
In order for $\partial_t \rho^{\infty}$ to be independent of $\epsilon$, $\textbf{u}^{\infty}$ must be incompressible, i.e. $\nabla\cdot\textbf{u}^{\infty}\!=\!0$. Incompressibility results in the sourceless advection equation
\begin{equation}
\partial_t \rho^{\infty} + (\textbf{u}^{\infty} \cdot \nabla) \rho^{\infty} = 0,
\end{equation}
implying that if density fluctuations $\rho^{\infty}$ are present in initial conditions, they are advected as a passive scalar, but no dynamical generation of $\rho^{\infty}$ is possible. This allows us to set $\rho^{\infty}\!=\!0$. Analogously, $\partial_t P$ corresponds to a sourceless advection equation for $P^{\infty}$, and we set $P^{\infty}\!=\!0$. As we will show later, density and pressure fluctuations are determined by the small amplitude compressible corrections to the $\infty$ component. The time derivative of $\textbf{u}$ gives
\begin{equation}
\begin{gathered}
\partial_t \textbf{u}^{\infty} = -(\textbf{u}^{\infty}\cdot\nabla)\textbf{u}^{\infty} + \frac{1}{\rho_0} ( \nabla\times\textbf{B}^{\infty} ) \times \textbf{B}^{\infty} + 
\\[5pt]
+ \frac{1}{\epsilon\rho_0}\bigl[ (\textbf{B}_0 \cdot \nabla) \textbf{B}^{\infty} - \nabla (\textbf{B}_0 \cdot \textbf{B}^{\infty}) \bigr],
\end{gathered}    
\end{equation}
which is independent of $\epsilon$ only if 
\begin{equation}
(\textbf{B}_0 \cdot \nabla) \textbf{B}^{\infty} = \nabla (\textbf{B}_0 \cdot \textbf{B}^{\infty}).
\end{equation}
The above condition implies $\partial_z B^{\infty}_x \!=\! \partial_x B^{\infty}_z$ and $\partial_z B^{\infty}_y \!=\! \partial_y B^{\infty}_z$, from which it follows
\begin{equation}
( \nabla\times\textbf{B}^{\infty} ) \times \textbf{B}^{\infty} = ( \nabla_{\perp}\times\textbf{B}_{\perp}^{\infty} ) \times \textbf{B}_{\perp}^{\infty},    
\end{equation}
where $\nabla_{\perp}\!=\!(\partial_x,\,\partial_y,\,0)$ and $\textbf{B}_{\perp}^{\infty}\!=\!(B_{x}^{\infty},\,B_{y}^{\infty},\,0)$. Finally, the time derivative of $\textbf{B}$ gives
\begin{equation}
\begin{gathered}
\partial_t \textbf{B}^{\infty} = -(\textbf{u}^{\infty} \cdot \nabla) \textbf{B}^{\infty} + (\textbf{B}^{\infty} \cdot \nabla) \textbf{u}^{\infty} + 
\\[5pt]
+\frac{1}{\epsilon} (\textbf{B}_0 \cdot \nabla) \textbf{u}^{\infty},
\end{gathered}    
\end{equation}
where the term containing $\epsilon$ vanishes only if
\begin{equation}
(\textbf{B}_0 \cdot \nabla) \textbf{u}^{\infty} = 0 \,\, \rightarrow \,\, \partial_z \textbf{u}^{\infty} = 0,    
\end{equation}
implying that $\textbf{u}^{\infty}$ is 2D. Combining all the above results, we obtain equations 
\begin{equation}
\nabla_{\perp} \cdot \textbf{u}_{\perp}^{\infty} = 0,
\label{INFrho}
\end{equation}

\begin{equation}
\rho_0 \bigl[ \partial_t  + \bigl( \textbf{u}_{\perp}^{\infty} \cdot \nabla_{\perp} \bigr) \bigr] \textbf{u}_{\perp}^{\infty} = \bigl( \nabla_{\perp} \times \textbf{B}_{\perp}^{\infty} \bigr) \times \textbf{B}_{\perp}^{\infty},
\label{INFu}
\end{equation}

\begin{equation}
\bigl[ \partial_t  + \bigl( \textbf{u}_{\perp}^{\infty} \cdot \nabla_{\perp} \bigr) \bigr] \textbf{B}_{\perp}^{\infty} = \bigl( \textbf{B}_{\perp}^{\infty} \cdot \nabla_{\perp} \bigr) \textbf{u}_{\perp}^{\infty},
\label{INFB}
\end{equation}
where $\textbf{u}_{\perp}^{\infty}\!=\!(u_{x}^{\infty},\,u_{y}^{\infty},\,0)$, plus the sourceless advection equation
\begin{equation}
\bigl[ \partial_t + (\textbf{u}_{\perp}^{\infty} \cdot \nabla_{\perp}) \bigr] u_z^{\infty} = 0,
\end{equation}
that allows us to set $u_z^{\infty}\!=\!0$, and equation
\begin{equation}
\bigl[ \partial_t + (\textbf{u}_{\perp}^{\infty} \cdot \nabla_{\perp}) \bigr] B_z^{\infty} = \bigl( \textbf{B}_{\perp}^{\infty} \cdot \nabla_{\perp} \bigr) u_z^{\infty},
\end{equation}
implying $B_z^{\infty}\!=\!0$, since $u_z^{\infty}\!=\!0$. The closed system (\ref{INFrho})-(\ref{INFB}) corresponds to 2D cold incompressible MHD\footnote{In the case of turbulence developing over a large-scale inhomogeneous background, equation $\nabla_{\perp} \cdot \textbf{u}_{\perp}^{\infty} \!=\! 0$ is replaced by a more general incompressibility condition \citep[see e.g.][]{zank2017theory}.}, describing the dynamics of the $\infty$ component. A major difference with respect to subsonic models is that pressure fluctuations do not affect the dynamics of the majority $\infty$ component.

We now introduce small amplitude corrections to the $\infty$ component, labeled as \enquote{$\star$}, of the form
\begin{equation}
\begin{gathered}
\rho = \rho_0 + \epsilon \rho^{\star},
\quad
\textbf{u} = \textbf{u}_{\perp}^{\infty} + \epsilon \textbf{u}^{\star},
\\[10pt]
P = P_0 + \epsilon P^{\star},
\quad
\textbf{B} = \textbf{B}_0 + \epsilon \textbf{B}_{\perp}^{\infty} + \epsilon^2 \textbf{B}^{\star}.
\end{gathered}
\end{equation}
Substituting the above expansion into equations (\ref{density}), (\ref{pressure})-(\ref{Tmomentum}), and subtracting (\ref{INFrho})-(\ref{INFB}), we obtain the leading-order equations for the $\star$ component
\begin{equation}
\begin{gathered}
\bigl[ \partial_t  + \bigl( \textbf{u}_{\perp}^{\infty} \cdot \nabla_{\perp} \bigr) \bigr] \rho^{\star} = -\rho_0 \bigl( \nabla\cdot\textbf{u}^{\star} \bigr),
\end{gathered}
\label{STARrho}
\end{equation}

\begin{equation}
\begin{gathered}
\rho_0 \bigl[ \partial_t  + \bigl( \textbf{u}_{\perp}^{\infty} \cdot \nabla_{\perp} \bigr) \bigr] \textbf{u}^{\star} + \rho_0 \bigl( \textbf{u}_{\perp}^{\star} \cdot \nabla_{\perp} \bigr) \textbf{u}_{\perp}^{\infty} = 
\\[5pt] 
= -\nabla P^{\star} -\frac{\rho^{\star}}{\rho_0} \bigl[ \bigl( \nabla_{\perp} \times \textbf{B}_{\perp}^{\infty} \bigr) \times \textbf{B}_{\perp}^{\infty} \bigl] + 
\\[5pt]
+ \bigl( \textbf{B}_{\perp}^{\infty} \cdot \nabla_{\perp} \bigr) \textbf{B}^{\star} + \bigl( \textbf{B}_{\perp}^{\star} \cdot \nabla_{\perp} \bigr) \textbf{B}_{\perp}^{\infty} +
\\[5pt]
- \nabla \bigl( \textbf{B}_{\perp}^{\infty} \cdot \textbf{B}_{\perp}^{\star} \bigr) + \frac{1}{\epsilon} \bigl[ \bigl( \textbf{B}_0 \cdot \nabla \bigr) \textbf{B}^{\star} - \nabla \bigl( \textbf{B}_0 \cdot \textbf{B}^{\star} \bigr) \bigr],
\end{gathered}
\label{STARu}
\end{equation}

\begin{equation}
\begin{gathered}
\bigl[ \partial_t  + \bigl( \textbf{u}_{\perp}^{\infty} \cdot \nabla_{\perp} \bigr) \bigr] P^{\star} = -\gamma P_0 \bigl( \nabla\cdot\textbf{u}^{\star} \bigr),
\end{gathered}
\label{STARP}
\end{equation}

\begin{equation}
\begin{gathered}
\bigl[ \partial_t  + \bigl( \textbf{u}_{\perp}^{\infty} \cdot \nabla_{\perp} \bigr) \bigr] \textbf{B}^{\star} = \nabla \times \bigl( \textbf{u}^{\star} \times \textbf{B}_{\perp}^{\infty} \bigr) + 
\\[5pt]
+ \bigl( \textbf{B}_{\perp}^{\star} \cdot \nabla_{\perp} \bigr) \textbf{u}_{\perp}^{\infty} + \frac{1}{\epsilon} \bigl[ \bigl( \textbf{B}_0 \cdot \nabla \bigr) \textbf{u}^{\star} - \textbf{B}_0 \bigl( \nabla \cdot \textbf{u}^{\star} \bigr) \bigr],
\end{gathered}
\label{STARB}
\end{equation}
where higher-order powers of $\epsilon$ have been neglected.

Equations (\ref{INFrho})-(\ref{INFB}) and (\ref{STARrho})-(\ref{STARB}) represent the transonic sub-Alfvénic limit of compressible MHD\footnote{Equations (\ref{INFrho})-(\ref{INFB}) and (\ref{STARrho})-(\ref{STARB}) can be rewritten in dimensional units by setting $\epsilon\!=\!M_A$ and by using the normalization described in \citet{zank2017theory}.}. The structure of these equations reveals several properties of transonic sub-Alfvénic turbulence. We see that the majority $\infty$ component is strictly 2D, incompressible, and contains only low frequencies. On the other hand, the minority $\star$ component is 3D, compressible, and contains both low and high frequencies, since time derivatives in equations (\ref{STARrho})-(\ref{STARB}) include both $\mathcal{O}(1)$ and $\mathcal{O}(1/\epsilon)$ terms, corresponding to time scales $T_S\!\sim\!T_T$ and $T_A$, respectively. The physical nature of the $\star$ component is better understood by noting that solutions to equations (\ref{STARrho})-(\ref{STARB}) in the absence of the $\infty$ component correspond to waves with dispersion relations
\begin{equation}
\omega^2_A = k^2 \frac{c_A^2}{\epsilon^2}\cos^2 \theta, 
\label{AW}
\end{equation}    

\begin{equation}
\omega^2_\pm = \frac{k^2}{2} \left[ c_M^2 \pm \sqrt{c_M^4 - 4\frac{c^2_S\,c^2_A}{\epsilon^2} \cos^2 \theta} \,\,\right],
\label{MS}
\end{equation}
where $k\!=\!|\textbf{k}|$, $\cos\theta\!=\!\textbf{k}\cdot\textbf{B}_0/k\,B_0$, and $c^2_M\!=\!c^2_S + c^2_A/\epsilon^2$. Subscripts \enquote{$A$}, \enquote{$-$} and \enquote{$+$} indicate AWs, SMs and FMs. Since $\epsilon\!\ll\!1$, we get
\begin{equation}
\omega^2_- \simeq k^2 c_S^2\cos^2 \theta, \qquad \omega^2_+ \simeq k^2 \frac{c_A^2}{\epsilon^2}.
\end{equation}
Hence, the solutions we find correspond to low frequency SMs, with $\omega_-\!\sim\!\mathcal{O}(1)$, and high frequency AWs and FMs, with $\omega_A\!\sim\!\omega_+\!\sim\!\mathcal{O}(1/\epsilon)$. In the presence of a nonzero $\infty$ component, the $\star$ component couples to the majority incompressible flow via the quasi-linear terms in equations (\ref{STARrho})-(\ref{STARB}). This coupling induces a frequency broadening around the eigenfrequencies of the above AWs, SMs and FMs solutions \citep[see e.g.][]{yuen2025temporal}. Thus, combining the $\infty$ and $\star$ components, we get a turbulent solution with a quasi-2D geometry, with most energy stored in low frequency 2D incompressible fluctuations, and a smaller amount of energy associated with frequency broadened waves, consisting of low frequency SMs, and high frequency AWs and FMs. Hence, our model shows that transonic sub-Alfvénic turbulence is effectively NI, with a 2D + slab geometry. All these properties predicted by our reduced model are in good agreement with our numerical results. Furthermore, we note that equations (\ref{INFrho})-(\ref{INFB}) are fully nonlinear, while equations (\ref{STARrho})-(\ref{STARB}) are quasi-linear, meaning that the $\infty$ component is the main actor in the turbulent cascade, as compared to the $\star$ component. This explains why the low frequency NWMs observed in our simulation exhibit a broadband $k_{\perp}$ energy distribution, while waves have relatively small wavenumbers, suggesting a weaker contribution to the cascade. 

The origin and properties of density and pressure fluctuations in transonic sub-Alfvénic turbulence can be inferred from equations (\ref{STARrho}) and (\ref{STARP}). We see that $\rho^{\star}$ and $P^{\star}$ are generated by the $\star$ component via compressible source terms proportional to $\nabla\cdot\textbf{u}^{\star}$. These source terms are $\mathcal{O}(1)$, since they contain contributions from both low frequency SMs and high frequency FMs. This is a major difference with respect to subsonic models, where density and pressure source terms are $\mathcal{O}(1/\epsilon)$, as both SMs and FMs are high frequency waves in the subsonic regime. After being generated by SMs and FMs, density and pressure fluctuations are advected by the incompressible turbulent flow $\textbf{u}^{\infty}_{\perp}$. As discussed in \citet{montgomery1987density} and \citet{zank2017theory}, this advection implies that $\rho^{\star}$ and $P^{\star}$ have spectral properties similar to $\textbf{u}^{\infty}_{\perp}$. This is consistent with the fact that $P_{\rho}$ in our simulation does not contain only SMs and FMs, but also NWMs corresponding to incompressible velocity fluctuations (see Figure~\ref{Prho}). Low frequency density NWMs potentially include zero frequency entropy modes \citep{zank2023linear,zank2024characterization}. An important relation between $\rho^{\star}$ and $P^{\star}$ is found by combining (\ref{STARrho}) and (\ref{STARP}), which gives
\begin{equation}
\bigl[ \partial_t  + \bigl( \textbf{u}_{\perp}^{\infty} \cdot \nabla_{\perp} \bigr) \bigr] \biggl( \frac{\rho^{\star}}{\rho_0} - \frac{P^{\star}}{\gamma P_0} \biggl) = 0,
\end{equation}
implying a linear correlation between pressure and density fluctuations along $\textbf{u}^{\infty}_{\perp}$ streamlines, known as \enquote{sound relation} 
\begin{equation}
\delta P^{\star} = \left( \gamma P_0/\rho_0 \right) \, \delta \rho^{\star} = c_S^2 \, \delta \rho^{\star}.
\label{sound_relation}
\end{equation}
This is another major difference with respect to subsonic models, where a low frequency $P^{\infty}$ is present, affecting the dynamics of the $\infty$ incompressible flow, and related to $\rho^{\star}$ by a \enquote{pseudosound relation} of the form $\delta P^{\infty} + \delta P^{\star} \!=\! c_S^2 \, \delta \rho^{\star}$. Equation~(\ref{sound_relation}) justifies the use of an isothermal closure in our simulation.

\section{Discussion and Conclusions}

In this Letter, we have investigated the properties of transonic sub-Alfvénic turbulence using a 3D MHD simulation initialized with near-Sun SW parameters, as observed by PSP. Our analysis shows that transonic turbulence is weakly compressible, mainly consisting of low frequency quasi-2D incompressible fluctuations, with minor compressible contributions corresponding to frequency broadened SMs and FMs. These numerical result are in good agreement with a new MHD model of Transonic sub-Alfvénic Turbulence (TsAT) that we have derived. Our model shows that in the $M_S\!\sim\!\mathcal{O}(1)$, $M_A\!\ll\!1$ regime, turbulent solutions to compressible MHD equations consist of a majority low frequency 2D incompressible component, plus a minority 3D compressible contribution containing low frequency SMs, and high frequency AWs and FMs. 

The main conclusion of our work is that transonic sub-Alfvénic turbulence lies in the NI realm, similarly to subsonic turbulence. This arguably unexpected result can be understood via the following intuitive argument. We can imagine that the formation of compressible fluctuations is determined by the competition between two mechanisms: on one side, turbulence tries to locally produce large density fluctuations on time scales $T_T$; on the other hand, compressible SMs and FMs tend to \enquote{diffuse away} density perturbations, propagating them throughout the plasma. The transonic sub-Alfvénic $M_S\!\sim\!\mathcal{O}(1)$, $M_A\!\ll\!1$ regime implies a temporal ordering where turbulence and SMs evolve on a similar time scale $T_T\!\sim\!T_S$, while FMs propagate on the much faster time scale $T_A$. Consequently, SMs are not fast enough to diffuse away density fluctuations produced by turbulence, but FMs are still able to counter the formation of compressible fluctuations, since $T_A\!\ll\!T_T$. This simple argument explains why transonic sub-Alfvénic turbulence is effectively NI. We note that despite its NI nature, transonic turbulence exhibits density fluctuations $\rho^{\star}/\rho_0\!\sim\!\mathcal{O}(\epsilon)\!\sim\!\mathcal{O}(M_A)$, and is thus more compressible than subsonic turbulence, where $\rho^{\star}/\rho_0\!\sim\!\mathcal{O}(\epsilon^2)\!\sim\!\mathcal{O}(M_A^2)$ \citep{zank1993nearly}. This different density ordering ultimately results from the fact that the subsonic $M_S,\,M_A\!\ll\!1$ regime implies $T_S,\,T_A\!\ll\!T_T$, meaning that both SMs and FMs contribute to the diffusion of density fluctuations, while in the transonic regime the diffusion process is less efficient, since it is mediated only by FMs. We expect that a strong compressible dynamics with $\delta \rho/\rho_0\!\sim\!\mathcal{O}(1)$ would occur only when turbulence is not just supersonic, but also super-Alfvénic, with $T_T\!\ll\!T_S,\,T_A$. 

To summarize, we have derived a new model that extends existing NI theories of SW turbulence, relaxing the subsonic assumption $M_S\!\ll\!1$. We have found that turbulence is NI and has a 2D + slab  geometry not only in the subsonic limit $M_S\!\ll\!1$, but also in the transonic regime $M_S\!\sim\!\mathcal{O}(1)$, as long as it remains sub-Alfvénic ($M_A\!\ll\!1$). Hence, our results suggest that turbulence retains a NI and 2D + slab nature also in the near-Sun SW, and possibly in the solar corona as well, where the $M_A\!\ll\!1$ condition is enforced by the strong local magnetic field, even if turbulence becomes transonic \citep{zank2018theory,adhikari2020solar,adhikari2022modeling}. Our results have potential implications for the theoretical and numerical modeling of near-Sun SW turbulence and its coupling with the solar corona. Our work is also potentially relevant for astrophysical applications where transonic sub-Alfvénic turbulence is expected to occur, as in the interstellar medium \citep{cho2002compressible,cho2003compressible}.

A few remarks are needed regarding the derivation of our TsAT model. We have looked for turbulent solutions to compressible MHD equations (\ref{density}), (\ref{pressure})-(\ref{Tmomentum}) in the $M_A\!=\!\epsilon\!\ll\!1$ regime, assuming that the real solution $\Psi$ can be approximated as $\Psi\!\simeq\!\Psi_0+\epsilon\,\Psi_1$, where $\Psi_0$ is a particular low frequency solution valid for $\epsilon\!\rightarrow\!0$, while $\Psi_1$ represents small amplitude corrections containing high frequency fluctuations. The choice of $\Psi_0$ is not unique in principle. Here, we have used Kreiss's theorem to obtain a $\Psi_0$ that does not contain AWs and magnetosonic modes, whose dispersion relations (\ref{AW})-(\ref{MS}) depend on $\epsilon$ (these waves are then reintroduced via $\Psi_1$). This choice is intentional but it is also possible to retain low frequency AWs and magnetosonic modes in $\Psi_0$. To do so, we write wavenumbers as $k\!=\!k_S+\epsilon\,k_L$, where $k_S$ represents small scales, while $k_L$ is associated with long-wavelength fluctuations. Consequently, MHD waves with dispersion relations (\ref{AW})-(\ref{MS}) can be split into two families, one including low frequency long-wavelength modes, with $k\!\simeq\!\epsilon\,k_L$ and $\omega(k_L)\!\sim\!\mathcal{O}(1)$, and the other containing high frequency short-wavelength fluctuations, with $k\!\simeq\!k_S$ and $\omega(k_S)\!\sim\!\mathcal{O}(1/\epsilon)$. Hence, by introducing two distinct length scales, the low frequency long-wavelength branch of MHD waves can be included into $\Psi_0$ using Kreiss's theorem, as discussed in \citet{zank1992equations}, where this approach has been used to derive \enquote{reduced MHD}. Moving low frequency MHD waves from $\Psi_1$ to $\Psi_0$ leads to slightly different equations than those we have derived, but with analogous physical properties. Another important caveat concerning our derivation is that the approach we have used to introduce corrections $\Psi_1$ to $\Psi_0$ is not a standard perturbation method, since it leads to the stiff equations (\ref{STARrho})-(\ref{STARB}), where $\mathcal{O}(1)$ and $\mathcal{O}(1/\epsilon)$ terms are mixed. This issue can be avoided using multiple time scale perturbation theory, producing a hierarchy of distinct equations for each order in $\epsilon$, as discussed in \citet{matthaeus1988nearly}. However, the method we have used has the advantage of giving a closed system of equations that is simple to interpret and provides important physical insights into the transonic sub-Alfvénic turbulent regime. 

Finally, we specify that our results concern inertial range turbulence at MHD scales, but kinetic scale dissipation has not been addressed. The higher level of compressibility characterizing transonic sub-Alfvénic turbulence may affect the anisotropic heating of ions and electrons via resonant wave-particle interactions \citep{cerri2021stochastic,verscharen2022electron,squire2022high,pezzini2024fully}, influencing dissipation and particle acceleration \citep{cerri2016subproton,hadid2017energy,zhdankin2021particle,arro2022spectral}. The kinetic properties of transonic sub-Alfvénic turbulence will be investigated in future works.


\vspace{10pt}

Research presented in this article was supported by the Laboratory Directed Research and Development program of Los Alamos National Laboratory under project number 20258088CT-SES.

This research used resources provided by the Los Alamos National Laboratory Institutional Computing Program, which is supported by the U.S. Department of Energy National Nuclear Security Administration under Contract No. 89233218CNA000001.

G.P.Z., L.Z., and L.A. acknowledge the partial support of a NASA PSP contract SV4-84017, an NSF EPSCoR RII-Track-1 Cooperative Agreement OIA-2148653, and a NASA IMAP contract 80GSFC19C0027.


\bibliography{TMHD}{}
\bibliographystyle{aasjournal}


\end{document}